\documentclass[prb,twocolumn]{revtex4-1} 
\usepackage{amsmath,amssymb,bm}
\usepackage{color}
\usepackage{fullpage}
\usepackage{hyperref}
\usepackage{natbib}
\usepackage{tikz}
\usetikzlibrary{decorations.pathreplacing,decorations.pathmorphing,shapes,calc,fadings,shadings,positioning}
\usetikzlibrary{decorations.markings}
\usepackage{braket}
\usepackage{graphicx}
\usepackage{subfigure}

\usepackage{xcolor}
\definecolor{myblue}{rgb}{49,130,189}

\newcommand{\tr}{\text{tr}}

\newcommand{\dg}{\dagger}
\begin{document}

\author{Dag-Vidar Bauer}
\author{J. O. Fj{\ae}restad}

\affiliation{Center for Quantum Spintronics, Department of Physics, Norwegian University of Science and Technology, NO-7491 Trondheim, Norway}

\title{Spin-wave study of entanglement and R\'{e}nyi entropy for coplanar and collinear magnetic orders in two-dimensional  quantum Heisenberg antiferromagnets}
\date{\today}

\begin{abstract} 
We use modified linear spin-wave theory (MLSWT) to study ground-state entanglement for a length-$L$ line subsystem 
in $L\times L$ square- and triangular-lattice  quantum Heisenberg antiferromagnets with coplanar spiral magnetic order with ordering vector $\bm{Q}=(q,q)$ and $N_G=3$ Goldstone modes, except if $q=\pi$ (collinear order, $N_G=2$). Generalizing earlier MLSWT results for $q=\pi$ to commensurate spiral order with $s\geq 3$ sublattices ($q=2\pi r/s$ with $r$ and $s$ coprime), we find analytically for large $L$ a universal and $n$-independent subleading term $(N_G/2)\ln L$ in the R\'{e}nyi entropy $S_n$, associated with $L^{1/2}$ scaling of $\lambda_0$ and $\lambda_{\pm q}$, with $\lambda_0\neq \lambda_{\pm q}$ for spiral order; here $\{\lambda_{k_y}\}$ are the $L$ mode occupation numbers of the entanglement Hamiltonian. The term $(3/2)\ln L$ in $S_n$ agrees with a nonlinear sigma model (NLSM) study of $s=3$ spiral order ($q=2\pi/3$). These and other properties of $S_n$ and  $\lambda_{k_y}$ are explored numerically for an anisotropic nearest-neighbor triangular-lattice model for which $q$ varies in the spiral phase.
\end{abstract}

\maketitle

\section{Introduction}
\label{introduction}

Entanglement, a concept originating in quantum information theory,\cite{nielsen-chuang} has turned out to be very useful for characterizing quantum many-body states,\cite{amico,jpmt-special-issue,laflorencie-review} not least universal ground state properties. The entanglement entropy and the more general R\'{e}nyi entropy have been particularly fruitful objects of study. These are measures of bipartite entanglement, defined in terms of the reduced density matrix $\rho_{\mathcal{A}} = \mbox{tr}_{\mathcal{B}} (|\Psi\rangle\langle \Psi|)$ for a subsystem $\mathcal{A}$, when the full system $\mathcal{A}\cup \mathcal{B}$ is in the pure quantum state $|\Psi\rangle$, taken to be the ground state in the following. The entanglement entropy is the von Neumann entropy of $\rho_{\mathcal{A}}$, 
\begin{equation}
S_{\rm{vN}} = -\tr_{\mathcal{A}} (\rho_{\mathcal{A}} \ln \rho_{\mathcal{A}}),
\end{equation}
and the R\'{e}nyi entropy, which depends on the R\'{e}nyi index $n$, is
\begin{equation}
S_n = -\frac{1}{n-1}\ln \tr_{\mathcal{A}} (\rho_{\mathcal{A}}^n),
\label{renyi-def}
\end{equation}
which reduces to the entanglement entropy in the limit $n\to 1$ ($S_1 \equiv \lim_{n\to 1}S_n = S_{\rm{vN}}$). 

In many classes of systems the leading term in the R\'{e}nyi entropy is found to scale linearly with the size of the boundary between $\mathcal{A}$ and $\mathcal{B}$,\cite{eisert,fr-survey} i.e. $\propto \ell^{d-1}$ in $d$ spatial dimensions for a connected subsystem $\mathcal{A}$ with characteristic linear size $\ell$. This "area law" originates in short-range entanglement across the boundary, implying a nonuniversal proportionality constant. \textit{Universal} properties of the ground state may however be reflected in the presence and form of \textit{subleading} terms. The first example of such a signature term was the topological entanglement entropy,\cite{TEE} a constant associated with topological order.\cite{wen-book} 

In the context of quantum antiferromagnets, topological order can occur for certain lattice quantum spin models with a gapped spin-liquid ground state.\cite{wen-book,vishwanath-review} However, for more typical models and interaction parameters, canonical examples being the nearest-neighbor Heisenberg model on the square and triangular lattices, the ground state has magnetic long-range order corresponding to the spontaneous breaking of continuous spin rotation symmetries. For the spin-1/2 model on the square lattice, following the observation of the area law in Quantum Monte Carlo (QMC) simulations of $S_2$ in Ref. \onlinecite{hastings2010}, Song et al.\cite{song2011} used modified linear spin wave theory (MLSWT) to study $S_n$ for an $(L/2)\times L$ cylinder subsystem in an $L\times L$ torus, and found an additive logarithmic correction $b_n \ln L$. From the absence of corners in the subsystem and the essentially $n$-independent value extracted for  $b_n$ ($b_1\approx 0.93$ and very close values for $n=2$-4), they suggested that the correction's origin was different than in conformally invariant critical systems. A log correction was also found in QMC calculations of $S_2$ in Ref. \onlinecite{kallin2011}, both for cylinder and square subsystems. For the latter, $b_2$ had the opposite sign and was much bigger than the expected log correction due to the corners. Ref. \onlinecite{kallin2011} proposed that the log correction was due to the N\'{e}el order, which in the finite-size system manifests itself in the low-lying "tower of states" (TOS) spectrum with level spacing $\propto L^{-d}$.\cite{anderson1952} 

By analyzing the O($N$) nonlinear sigma model (NLSM) and a model of two coupled O($N$) rotors,  
Metlitski and Grover\cite{MG} (MG) argued that in a system with O($N$) $\to$ O($N-1$) continuous symmetry breaking, $S_n$ acquires an $n$-independent subleading term $b\ln(\rho_s L^{d-1}/c)$, where $\rho_s$ is the spin stiffness and $c$ is the spin-wave velocity. MG emphasized the importance of both the spin-wave gap $c/L$ and the tower-of-states spectrum with level spacing $c^2/(\rho_s L^d)$ for this result, the argument of the log being the ratio of these energy scales. MG found that $b=(N-1)/2$ where $N-1$ is the number of Goldstone modes $N_G$, giving a universal term $b(d-1)\ln L$, and suggested that a universal logarithmic correction with coefficient $(N_G/2)(d-1)$ would be present also in other models with a $\bm{k}$-linear dispersion of Goldstone modes. MG also studied the spectrum of the entanglement Hamiltonian $H_E$ defined via $\rho_{\mathcal{A}}\propto\exp(-H_E)$ and found that it had the same TOS form at low "energies" as the Hamiltonian.  

Following these early developments, many authors have investigated MG's prediction of a logarithmic correction proportional to the number of Goldstone modes.\cite{v2} For lattice spin models with collinear magnetic order, these include QMC\cite{humeniuk2012,helmes2014,kulchytskyy2015,luitz2015} and MLSWT\cite{luitz2015,laflorencie2015,frerot2015} studies of the Heisenberg antiferromagnet on the square\cite{humeniuk2012,helmes2014,luitz2015,laflorencie2015,frerot2015} and cubic\cite{frerot2015} lattice, the XY antiferromagnet on the square lattice,\cite{kulchytskyy2015} and the XY ferromagnet on the square\cite{luitz2015,frerot2015} and cubic\cite{frerot2015} lattice. Here the respective ground states have SU(2) symmetry breaking with $N_G=2$ for the Heisenberg models and U(1) symmetry breaking ($N_G=1$) for the XY models. Independence of farther-neighbor interactions within the same phase (i.e. universality) and of the R\'{e}nyi index have also been explored.\cite{luitz2015,laflorencie2015} Most of the works found results consistent with the MG prediction $b=N_G/2$. An exception is "early" QMC studies,\cite{kallin2011,humeniuk2012,helmes2014} where the deviations in the extracted value of $b$ have been primarily attributed to the small system sizes that are accessible, but limitations to finite temperature and differences in the definitions  \cite{humeniuk2012} used for the boundary length have also been noted, as well as a lower "signal-to-noise" ratio for relevant quantities in the Heisenberg vs. the XY model.\cite{kulchytskyy2015} Other subleading terms in the R\'{e}nyi entropy have also been investigated, including  MG's prediction of a universal ''constant'' $\gamma_n^{\rm{ord}}$.\cite{kulchytskyy2015,laflorencie2015} 

Moving on to noncollinear magnetic order, a central example is the 3-sublattice coplanar order with 120 degrees between the ordering directions of neighboring spins, as found e.g. 
in the spin-1/2 Heisenberg antiferromagnet on the triangular lattice.\cite{trlattorder} Compared to the square-lattice model, which has 2-sublattice collinear order, the energies of the tower states are $\propto S(S+1)/L^2$ for both models ($S$ is the total spin quantum number), but the degeneracies differ: $2S+1$ for the 2-sublattice order and $(2S+1)^2$ for the 3-sublattice order.\cite{bernu1994,degeneracy,splitting}

So far, relatively little work has been done to study the entanglement spectrum and R\'{e}nyi entropy for models with noncollinear  magnetic order. Part of the reason may be that the QMC method extensively used for collinear order\cite{hastings2010,kallin2011,humeniuk2012,helmes2014,kulchytskyy2015,luitz2015} is not applicable to generic models with frustrated interactions due to the sign problem.\cite{sign-problem} Therefore other methods become all the more valuable. Kolley et al.\cite{kolley2013} used the density-matrix renormalization group (DMRG) method to study the entanglement spectrum for two models with the 3-sublattice order: the spin-1/2 Heisenberg model with an additional ferromagnetic next-nearest neighbor interaction $J_2=-J_1$ on the triangular and the kagome lattice. Similar to the collinear case, a correspondence was found between the low-energy entanglement spectrum and the low-energy spectrum of the Hamiltonian. Rademaker,\cite{rademaker2015} extending the NLSM approach of Ref. \onlinecite{MG} to triangular-lattice Heisenberg antiferromagnets with 3-sublattice order, also obtained such a correspondence, and in addition found a universal logarithmic correction in the R\'{e}nyi entropy with coefficient $b=3/2$, consistent with $N_G=3$ Goldstone modes. He furthermore obtained the dependence of the entanglement spectrum and the R\'{e}nyi entropy on the anisotropic spin stiffnesses and spin-wave velocities.

Motivated by the previous MLSWT studies of the R\'{e}nyi entropy for antiferromagnets with collinear order,\cite{song2011,luitz2015,laflorencie2015,frerot2015} here we generalize the MLSWT approach to Heisenberg exchange interactions with a more general Fourier transform $J(\bm{k})$, such that magnetic order in the ground state is generally coplanar, with collinear order as a special case. Although our theory is formulated on a square lattice, it may by suitable choice of $J(\bm{k})$ also describe Heisenberg models on the triangular lattice. We consider the simplest corner-free subsystem that also allows for an extraction of the universal logarithmic correction to the R\'{e}nyi entropy due to magnetic order, namely a straight one-dimensional line of length $L$ that wraps around an $L\times L$ torus.\cite{luitz2015} This choice of subsystem also makes the problem analytically solvable,\cite{luitz2015} which enables more insight into the solution than a purely numerical calculation would. 

This paper is organized as follows. Sec. \ref{theory} presents the general theory. It is applied to 
a triangular-lattice model with anisotropic nearest-neighbor interactions in Sec. \ref{results}. A summary and discussion is given in Sec. \ref{discussion}.

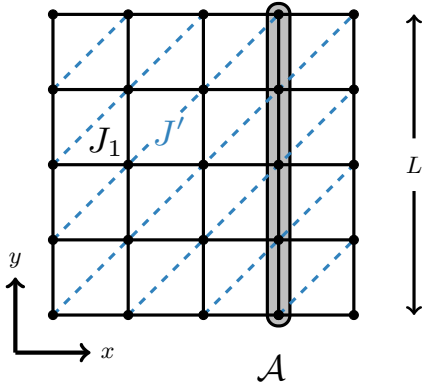
\begin{figure}[h]
\centering


\begin{tikzpicture}[scale=1]
\pgfmathsetmacro\si{sin(90)}
\pgfmathsetmacro\co{cos(90)}
\definecolor{blue2}{RGB}{49,130,189}


\draw[fill=gray!50,very thick,rounded corners] (1-0.15,0-0.145)--(1+0.15,0-0.145)--(1+0.15,4+0.145)--(1-0.15,4+0.145)--cycle;

\foreach \i in {0,1,...,4} {
\draw[very thick] ($ (-2,0) + (0,\i) $) -- ($ (2,\i) + (0,0) $);
}
\foreach \i in {1,...,5} {
\draw[very thick] ($ (-3,0) + (\i,0) $) -- ($ (-3,4) + (\i,0) $);
}

 \node[scale = 1.5] at (0.9,-0.7){$\mathcal{A}$};
  \node[scale = 1.5] at (-1.3,2.3){$J_1$};
%
    \draw[->,ultra thick] (-2.5,-0.5)--(-1.5,-0.5) node[right]{$x$};
    \draw[->,ultra thick] (-2.5,-0.5)--(-2.5,0.5) node[above]{$y$};


\draw[very thick, dashed, blue2] ($ (-2,3) $) -- ($(-1,4) $);
\draw[very thick, dashed, blue2] ($ (-2,2) $) -- ($(0,4) $);
\draw[very thick, dashed, blue2] ($ (-2,1) $) -- ($(-0.7,2.3) $);
\node[scale = 1.5, blue2] at (-0.45,2.45){$J'̈́$};
\draw[very thick, dashed, blue2] ($ (-0.3,2.7) $) -- ($(1,4) $);
\draw[very thick, dashed, blue2] ($ (-2,0) $) -- ($(2,4) $);
\draw[very thick, dashed, blue2] ($ (-1,0) $) -- ($(2,3) $);
\draw[very thick, dashed, blue2] ($ (0,0) $) -- ($(2,2) $);
\draw[very thick, dashed, blue2] ($ (1,0) $) -- ($(2,1) $);

\draw[arrows=<-,very thick] (2.8,0)--(2.8,2-0.35);
\node at (2.8,2){$L$};
\draw[arrows=->,very thick] (2.8,2+0.35)--(2.8,4.0);

	\foreach \x in {,-2,...,2}{
		\draw[black, fill] (\x,0) circle (0.06) ;}
	\foreach \x in {-2,...,2}{
		\draw[black, fill] (\x,1) circle (0.06) ;}
	\foreach \x in {-2,...,2}{
		\draw[black, fill] (\x,2) circle (0.06) ;}
	\foreach \x in {-2,...,2}{
		\draw[black, fill] (\x,3) circle (0.06) ;}        
	\foreach \x in {-2,...,2}{
		\draw[black, fill] (\x,4) circle (0.06) ;}   
\end{tikzpicture}

\caption{Square lattice with $N=L \times L$ sites (black dots). The Heisenberg interactions in the anisotropic triangular-lattice model are also shown ($J_1$ on the solid black bonds and $J'$ on the dashed blue bonds; note that the $J'$ bonds only involve one of the two diagonals on each square plaquette). Periodic boundary conditions are assumed in both directions, giving the system the topology of a torus. The subsystem $\mathcal{A}$ consists of all sites inside the thick shaded line that wraps around the 
torus along the $y$ direction at some fixed $x_i$. }
\label{square-triangle}
\end{figure}


\section{Theory}
\label{theory}


\subsection{Coplanar or collinear order in a classical Heisenberg model}
\label{classical}

We consider a two-dimensional square lattice of $N=L^2$ sites with periodic boundary conditions in both directions. The lattice sites, at positions $\bm{r}_i=(x_i,y_i)$, are occupied by classical spins $\bm{S}_i$ ($i=1,\ldots,N$) which interact via translationally invariant exchange interactions $J(\bm{r}_i-\bm{r}_j)\equiv J_{ij}=J_{ji}$, such that the Hamiltonian is given by the Heisenberg model 
\begin{equation}
H = \frac{1}{2}\sum_{i,j}J_{ij} \bm{S}_i \cdot \bm{S}_j.
\label{Heis-model}
\end{equation}
Fig. \ref{square-triangle} shows the $J_{ij}$ in a specific model to be considered later; note that this is equivalent to a model on the triangular lattice.

Introducing Fourier transforms\cite{inverse-transforms}
\begin{eqnarray}
\bm{S}(\bm{k}) &=& \frac{1}{\sqrt{N}}\sum_{j} e^{-i\bm{k}\cdot\bm{r}_j}\bm{S}_j, \\
J(\bm{k}) &=& \sum_{j}J_{ij}e^{-i\bm{k}\cdot (\bm{r}_i-\bm{r}_j)},
\end{eqnarray}
where $J(\bm{k})$ is real and even in $\bm{k}$, gives  
\begin{equation}
H = \frac{1}{2}\sum_{\bm{k}}J(\bm{k})\bm{S}(\bm{k})\cdot \bm{S}(-\bm{k}) 
\end{equation}
where the sum is over the first Brillouin zone. Invoking Parseval's theorem $\sum_{\bm{k}} \bm{S}(\bm{k})\cdot \bm{S}(-\bm{k}) = NS^2$ ($S$ is the spin length), it follows that the energy is minimized by putting all weight into the $\bm{k}$-vector(s) that minimize $J(\bm{k})$. This gives the classical ground state energy $E_{0,\rm{classical}} = \frac{1}{2}J_{\rm{min}}NS^2$ where $J_{\rm{min}}$ is the global minimum of $J(\bm{k})$. 

We will assume that the minima of $J(\bm{k})$ in the first Brillouin zone satisfy certain properties. We now discuss these and the types of magnetic ordering patterns that arise as a consequence. 

$J(\bm{k})$ has one or two minima. If $J(\bm{k})$ has one minimum, it occurs at $\bm{Q}=(\pi,\pi)$. This gives ground state spin configurations with $\bm{S}_i=S\bm{n}_1(-1)^{x_i+y_i}$, which is a collinear order along $\bm{n}_1$, a unit vector which labels/distinguishes different configurations. If instead $J(\bm{k})$ has two minima, they occur at $\pm \bm{Q}$, with $\bm{Q}=(q,q)$ ($\pi/2<q<\pi)$. (Note that this e.g. excludes the case of having two minima at $(\pi,0)$ and $(0,\pi)$.) Then a ground state spin configuration can be labeled by two orthogonal unit vectors $\bm{n}_1$ and $\bm{n}_2$. They span a plane within which all spins lie, while the spin directions within the plane describe a spiral structure determined by $\bm{Q}$:\cite{classical}
\begin{equation}
\bm{S}_i = S[\bm{n}_1 \cos(\bm{Q}\cdot\bm{r}_i) + \bm{n}_2 \sin(\bm{Q}\cdot \bm{r}_i)].
\label{classical-order}
\end{equation}
We will refer to this magnetic order as coplanar spiral (spiral for short). Note that the collinear order along $\bm{n}_1$ described above, corresponding to $\bm{Q}=(\pi,\pi)$, is also captured by Eq. (\ref{classical-order}). For both the spiral and collinear orders, $\bm{Q}$ will be referred to as the ordering vector.

\subsection{Linear spin wave theory}

We will now generalize to quantum spins of spin quantum number $S$ and study the corresponding quantum Heisenberg model using linear spin wave theory (LSWT). For concreteness we assume that the order (\ref{classical-order}) of the classical model is in the $zx$ plane with $\mathbf{n}_1=\hat{\mathbf{z}}$ and $\mathbf{n}_2=\hat{\mathbf{x}}$. To enable a $1/S$-expansion, we introduce rotated spin components $S_i^{\tilde{z}}$, $S_i^{\tilde{x}}$ where the local $\tilde{z}$ axis is chosen to coincide with the classical ordering direction of $\bm{S}_i$. Thus 
\begin{subequations}
\begin{eqnarray}
S^z_i &=& S^{\tilde{z}}_i\cos \theta_i - S^{\tilde{x}}_i \sin \theta_i, \\
S^x_i &=& S^{\tilde{z}}_i\sin \theta_i + S^{\tilde{x}}_i \cos \theta_i
\end{eqnarray}
\end{subequations}
(and $S^y_i = S^{\tilde{y}}_i$), where $\theta_i = \mathbf{Q}\cdot \mathbf{r}_i$ is the angle between the ordering direction of $\mathbf{S}_i$ and the $z$ axis. This gives
\begin{eqnarray}
\lefteqn{H = \frac{1}{2}\sum_{i,j}J_{ij}\Big[\cos(\theta_i-\theta_j)(S_i^{\tilde{x}} S_j^{\tilde{x}} + S_i^{\tilde{z}} S_j^{\tilde{z}}) \;\;+} \nonumber \\ & & \hspace{-0.6cm}\sin(\theta_i-\theta_j)(S_i^{\tilde{z}} S_j^{\tilde{x}} - S_i^{\tilde{x}} S_j^{\tilde{z}}) + S_i^{\tilde{y}} S_j^{\tilde{y}}\Big] 
- h\sum_i  S_i^{\tilde{z}}.
\end{eqnarray}
For later use we have here added by hand a term $\propto h$, where $h > 0$ is a fictitious local magnetic field along the $\tilde{z}$ direction. 

Next, we invoke the Holstein-Primakoff (HP) representation for the spin components,
\begin{subequations}
\begin{align}
& S_i^+ = \sqrt{2S-n_i}~b_i \\
& S_i^- = b_i^\dg\sqrt{2S-n_i}  \\
& S_i^{\tilde{z}} = S - n_i
\end{align}
\end{subequations}
where $S^{\pm} \equiv S^{\tilde{x}}\pm i S^{\tilde{y}}$, $n_i = b_i^\dg b_i$, and $b_i^{\dagger}$, $b_i$ are canonical bosonic creation and annihilation operators. In LSWT we ignore terms in $H$ of higher order than quadratic in HP bosons, which amounts to truncating the square root expansion to lowest order in $1/S$. The Hamiltonian then becomes (we omit constants in $H$ in the following) 
\begin{eqnarray}
H &=& \frac{S}{2}\sum_{i,j}J_{ij} \Big\{\frac{1}{2}\left[\cos(\theta_i-\theta_j)+1\right](b_i b_j^\dg + b_i^\dg b_j) \nonumber \\
&+& \frac{1}{2}\left[\cos(\theta_i-\theta_j)-1\right]\left( b_i b_j + b_i^\dg b_j^\dg\right) \nonumber \\
&-& \cos(\theta_i-\theta_j)(n_i+n_j) \Big\}  + h\sum_i n_i.
\end{eqnarray}
Introducing canonical boson operators $b_{\bm{k}}$ via the Fourier transform $b_{\bm{k}} = \frac{1}{\sqrt{N}}\sum_j b_j e^{-i \bm{k} \cdot \bm{r}_j}$ gives
\begin{equation}
\hspace{-0.5cm}H = \frac{1}{2}\sum_{\bm{k}} \left[ A_{\bm{k}} (b_{\bm{k}}^\dg b_{\bm{k}} + b_{-\bm{k}} b_{-\bm{k}}^\dg) + B_{\bm{k}} (b_{-\bm{k}} b_{\bm{k}} + \mbox{h.c.})\right]
\end{equation}
where
\begin{subequations}
\begin{eqnarray}
A_{\bm{k}} &=& \frac{S}{2} \left[ 
\frac{1}{2}(J(\bm{k}-\bm{Q}) + J(\bm{k}+\bm{Q})) + J(\bm{k})-2J(\bm{Q})
\right] \nonumber \\ &+& h, \label{Ak} \\ B_{\bm{k}} &=& \frac{S}{2}\left[\frac{1}{2}(J(\bm{k}-\bm{Q}) + J(\bm{k}+\bm{Q})) - J(\bm{k}) \right]. \label{Bk}
\end{eqnarray}
\label{AB}
\end{subequations}
The Hamiltonian can be put in diagonal form by doing the Bogoliubov transformation 
\begin{equation}
b_{\bm{k}} = \cosh{\zeta_{\bm{k}}} \;\alpha_{\bm{k}} - \sinh{\zeta_{\bm{k}}} \;\alpha_{-\bm{k}}^\dagger
\end{equation}
with $\zeta_{\bm{k}}$ real and even. By choosing
\begin{equation}
\tanh 2\zeta_{\bm{k}} = \frac{B_{\bm{k}}}{A_{\bm{k}}}
\end{equation}
	the off-diagonal terms in $H$ vanish, i.e. 
\begin{align} H  = \sum_{\bm{k}} ~\omega_{\bm{k}} \alpha_{\bm{k}}^\dg \alpha_{\bm{k}},
 \end{align}
where the spin-wave dispersion $\omega_{\bm{k}}$ is given by
\begin{widetext}
\begin{equation}
\omega_{\bm{k}} = \sqrt{A_{\bm{k}}^2 - B_{\bm{k}}^2} = S \sqrt{\left[\frac{1}{2}(J(\bm{k}-\bm{Q}) + J(\bm{k}+\bm{Q})) - J(\bm{Q}) + \frac{h}{S}\right]\left[J(\bm{k})-J(\bm{Q}) + \frac{h}{S}\right]}.
\label{dispersion}
\end{equation}
\end{widetext}

The $N$- and $h$-dependent sublattice magnetization $m(N,h) \equiv N^{-1}\sum_i \langle S^{\tilde{z}}_i\rangle$ can be written ($h>0$)
\begin{equation}
m(N,h) = S + \frac{1}{2}-\frac{1}{2N}\sum_{\bm{k}} \frac{A_{\bm{k}}}{\omega_{\bm{k}}}. 
\end{equation}
In the thermodynamic limit, spin rotation symmetry is broken in the ordered phase, giving
\begin{equation}
\langle \bm{S}_i\rangle = m_{\rm{AF}}[\bm{n}_1 \cos(\bm{Q}\cdot\bm{r}_i) + \bm{n}_2 \sin(\bm{Q}\cdot \bm{r}_i)]
\end{equation}
(cf. Eq. (\ref{classical-order})), where the order parameter $m_{\rm{AF}}$ is 
\begin{eqnarray}
m_{\rm{AF}} & \equiv & \lim_{h\to 0}\lim_{N\to\infty}m(N,h) \nonumber \\ &=& 
S + \frac{1}{2} - \frac{1}{8\pi^2}\int d^2 k\; \frac{A_{\bm{k}}(h=0)}{\omega_{\bm{k}}(h=0)}.
\end{eqnarray}
It is seen from Eq. (\ref{dispersion}) that in the limit $h\to 0$, $\omega_{\bm{k}}$ vanishes at $\bm{k}=0$ and $\pm \bm{Q}$, signifying the $N_G$ Goldstone modes in the ordered phase. The set of $N_G$ inequivalent wavevectors  where this vanishing occurs will be referred to as $G$, i.e. $G=\{\bm{0},\bm{Q}\}$ for collinear order ($N_G=2$) and 
$G=\{\bm{0},\pm \bm{Q}\}$ for spiral order ($N_G=3$).

\subsection{Modified linear spin-wave theory for   finite-size systems}
\label{local-field}

In this work we wish to analyze a system of finite size $N$. Then there is no broken spin-rotation symmetry, so $\langle S^{\tilde{z}}_i\rangle$, and thus also $m(N,h)$, should be 0. This can be achieved by tuning the value of $h$, which thus becomes a function of $N$. This defines the modified linear spin-wave theory (MLSWT).\cite{song2011} Following Ref. \onlinecite{laflorencie2015} we define 
\begin{equation}
	m^*(N,h) = S+\frac{1}{2} -\frac{1}{2N}\sum_{\bm{k}\notin G}\frac{A_{\bm{k}}}{\omega_{\bm{k}}}.
    \label{mstar-def}
\end{equation}  
Requiring $m(N,h)=0$ gives the equation for $h$:
\begin{equation} 
m^*(N,h) = \frac{1}{2N}\sum_{\bm{k}\in G}\frac{A_{\bm{k}}}{\omega_{\bm{k}}}.
\label{h-eq-1}
\end{equation}
Here we assume that $\bm{k}$ can indeed equal $\bm{Q}$, i.e. that $q$ coincides with a valid wavevector $2\pi m_y/L$ where $m_y$ is an integer. This gives a magnetic order commensurate with the lattice, with the number $s$ of magnetic sublattices given via $m_y/L=r/s$, where the positive integers $r$ and $s$ are coprime (thus $L$ is always a multiple of $s$, so the order is not frustrated by the periodic boundary conditions). Using that $A_{\bm{k}}$ and $\omega_{\bm{k}}$ are even in $\bm{k}$, and introducing 
\begin{subequations}
\begin{eqnarray}
\Upsilon_{\bm{0}} &=& \sqrt{J(\bm{0})-J(\bm{Q})},\\
 \Upsilon_{\bm{Q}} &=& \sqrt{\frac{1}{2}(J(\bm{0})+J(2\bm{Q}))-J(\bm{Q})},
\end{eqnarray}
\end{subequations}
Eq. (\ref{h-eq-1}) can be written
\begin{eqnarray}
m^*(N,h)
&=& \frac{1}{N}\frac{\frac{1}{4}\Upsilon_{\bm{0}}^{2}+\frac{h}{2S}}{\sqrt{\frac{h}{S}\left(\Upsilon_{\bm{0}}^{2}+\frac{h}{S}\right)}} \nonumber \\ & & \hspace{-1.5cm}+\; (N_G-1) \frac{1}{N}\frac{\frac{1}{4}\Upsilon_{\bm{Q}}^{2}+\frac{h}{2S}}{\sqrt{\frac{h}{S}\left(\Upsilon_{\bm{Q}}^{2}+\frac{h}{S}\right)}}.
\label{h-eq-2}
\end{eqnarray}
The collinear phase has $N_G=2$ and $J(2\bm{Q})=J(\bm{0})$, so the two terms on the right-hand side are equal. This gives
\cite{laflorencie2015}
\begin{equation}
h = \frac{S\Upsilon_{\bm{0}}^{2}}{2}\left( \frac{1}{\sqrt{1-\frac{1}{(Nm^*(N,h))^2}}}-1\right). 
\end{equation}
For a general (not necessarily collinear) phase, Eq. (\ref{h-eq-2}) can only be solved
approximately. Expanding the right-hand side to zeroth order in the small quantities $h\Upsilon_{\bm{0},\bm{Q}}^{-2}/S$ gives 
\begin{equation}
h \approx \frac{S}{16 (N m^*(N,h))^2}\left(\Upsilon_{\bm{0}} + (N_G-1)\Upsilon_{\bm{Q}}\right)^2.
\label{h-leading-iteration}
\end{equation}
To calculate $h$ numerically, we have iterated Eqs. (\ref{h-leading-iteration}) and (\ref{mstar-def}) until convergence is reached. We find the approximation (\ref{h-leading-iteration}) to be sufficiently accurate for our calculations (the leading correction is $\propto (N m^*(N,h))^{-4}$).  

\subsection{Reduced density matrix and R\'{e}nyi entropy} 
\label{RDM}

As density matrix eigenvalues are real, nonnegative, and sum to 1, the reduced density matrix $\rho_{\mathcal{A}}$ for an arbitrary subsystem $\mathcal{A}$ can be parameterized as $\rho_{\mathcal{A}}=e^{-H_E}/\text{Tr}(e^{-H_E})$, where the ``entanglement Hamiltonian'' operator $H_E$ for the subsystem is hermitian. As the Hamiltonian $H$ is quadratic in boson operators, a (generalized) Wick's theorem\cite{wick} holds, so any correlation function $\langle \mathcal{O}\rangle$ can be expressed in terms of two-point correlation functions of boson operators. Specializing to operators $\mathcal O$ involving subsystem $\mathcal{A}$ only, in which case $\langle \mathcal O \rangle = \mbox{tr}_{\mathcal{A}} (\rho_{\mathcal{A}} \mathcal O)$, it follows that $H_E$ is also quadratic, and it can be determined entirely from two-point correlators inside $\mathcal{A}$.\cite{peschel} This involves a diagonalization procedure that for a subsystem of general shape must be done numerically. In contrast, we will here consider a subsystem consisting of all $L$ sites with a fixed value of $x$ (see Fig. \ref{square-triangle}). The translational invariance in the $y$ direction can then be exploited to simplify the analysis, leading to exact expressions for $H_E$ and other quantities of interest. 

To this end, we write $b_j=b_{x,y}$ and define new canonical boson operators
\begin{eqnarray}
b_{x,k_y} =  \frac{1}{\sqrt{L}}\sum_{y} e^{-i k_y y} b_{x,y}.
\end{eqnarray}
They have the two-point functions
\begin{subequations}
\begin{eqnarray}
\langle b_{x,k_y}b^{\dagger}_{x,k'_y}\rangle &=& \frac{1}{2} \delta_{k_y,k'_y} \bigg(\mathcal C(k_y) + 1\bigg), \label{2p1}\\
\langle b_{x,k_y}b_{x,k'_y}\rangle &=& - \frac{1}{2}\delta_{k_y,-k'_y} \; \mathcal{S}(k_y), \label{2p2}
\end{eqnarray}
\end{subequations}
where  
\begin{subequations}
\begin{eqnarray}
\mathcal C(k_y) &=& \overline{\cosh 2\zeta}(k_y), \\
\mathcal S(k_y) &=& \overline{\sinh 2\zeta}(k_y),
\end{eqnarray}
\end{subequations}
where the horizontal line denotes an average over $k_x$, 
\begin{equation}
\overline{f}(k_y) \equiv \frac{1}{L}\sum_{k_x}f(k_x,k_y).
\end{equation}
The Kronecker deltas in (\ref{2p1})-(\ref{2p2}), a consequence of the translational invariance in the $y$ direction, suggest defining the $2\times 2$ ``correlation matrix'' 
\begin{eqnarray}
M_{x,k_y} & \equiv & \left\langle \left( \begin{array}{c} b_{x,k_y} \\ b_{x,-k_y}^{\dagger} \end{array} \right)\left( \begin{array}{cc} b_{x,k_y}^{\dagger} & b_{x,-k_y} \end{array} \right)\right\rangle \nonumber \\ &=& 
\frac{1}{2}\left( \begin{array}{cc} 
\mathcal C(k_y)  + 1 & 
- \mathcal S(k_y) \\ 
- \mathcal S(k_y) & 
\mathcal C(k_y) - 1 
\end{array}\right).
\end{eqnarray}
A Bogoliubov transformation $b_{x,k_y} = \cosh \eta_{k_y}\beta_{x,k_y} - \sinh \eta_{k_y}\beta_{x,-k_y}^{\dagger}$ ($\eta_{k_y}$ real and even) implies that 
\begin{equation}
P_{k_y}^{-1} M_{x,k_y} (P_{k_y}^{\dagger})^{-1}=
\left( \begin{array}{cc} 
\langle \beta_{x,k_y} \beta^{\dagger}_{x,k_y} \rangle  & 
\langle \beta_{x,k_y} \beta_{x,-k_y} \rangle \\ 
\langle \beta^{\dagger}_{x,-k_y} \beta^{\dagger}_{x,k_y}\rangle & 
\langle \beta^{\dagger}_{x,-k_y}\beta_{x,-k_y}\rangle 
\end{array} \right)
\label{corrdiag}
\end{equation}
where
\begin{equation}
P_{k_y} = \left( \begin{array}{cc} \cosh \eta_{k_y} & -\sinh \eta_{k_y} \\ -\sinh \eta_{k_y} & \cosh \eta_{k_y} \end{array}\right). 
\end{equation}
The bosonic operators $\beta_{k_y}$ are chosen to be the set in terms of which $H_E$ is diagonal, i.e. 
\begin{equation}
H_E = \sum_{k_y}\varepsilon_{k_y}\beta_{x,k_y}^{\dagger}\beta_{x,k_y}.
\label{H-E}
\end{equation}
Then the off-diagonal terms in (\ref{corrdiag}) vanish, which leads to the condition
\begin{equation}
\tanh 2\eta_{k_y} = \frac{\mathcal S(k_y)}{\mathcal C(k_y)}.
\end{equation}
The diagonal terms are $1+\lambda_{k_y}$ and $\lambda_{k_y}$, where
$\lambda_{k_y}$ is the boson occupation number for the mode labeled by wavevector $k_y$:
\begin{eqnarray}
\lefteqn{\hspace{-0.5cm}\lambda_{k_y} = \langle \beta^{\dagger}_{x,k_y}\beta_{x,k_y}\rangle = \frac{1}{2}\left[\sqrt{\mathcal C^2(k_y) - \mathcal S^2(k_y)} - 1\right]} \nonumber \\ & & \hspace{-0.5cm}= 
\frac{1}{2}\left[
\sqrt{\left(\frac{1}{L}\sum_{k_x}\frac{A_{\bm{k}}}{\omega_{\bm{k}}}\right)^2 - \left(\frac{1}{L}\sum_{k_x}\frac{B_{\bm{k}}}{\omega_{\bm{k}}}\right)^2} - 1\right].
\label{lambdaky-1}
\end{eqnarray}
The occupation numbers for $k_y=0$ and $k_y=\pm q$ will be of particular interest in the subsequent analysis. For the collinear phase it can be shown\cite{proof-symm-collinear} that
\begin{equation}
\lambda_0 = \lambda_{q}.
\label{symm-collinear}
\end{equation}

The mode energy $\varepsilon_{k_y}$ in $H_E$ is related to $\lambda_{k_y}$ via the Bose-Einstein distribution, $\lambda_{k_y}=[\exp(\varepsilon_{k_y})-1]^{-1}$. Finally, the R\'{e}nyi entropy (\ref{renyi-def}) can be expressed in terms of either set of quantities. Using the former set gives  
\begin{equation}\label{renyi-lambda}
S_n = \frac{1}{n-1}\sum_{k_y}\ln\left[(1+\lambda_{k_y})^n - \lambda_{k_y}^n\right]. 
\end{equation}

\subsection{Analytical approximations for large $L$}
\label{large-L}

In the limit of large $N$, when $m^*(N,h)\to m_{\rm{AF}}$, (\ref{h-leading-iteration}) simplifies to
\begin{equation}
h = \frac{1}{16N^2}\frac{S}{ m_{AF}^2}\left(\Upsilon_{\bm{0}} + (N_G-1)\Upsilon_{\bm{Q}}\right)^2
\label{h-leading}
\end{equation}
(making the gap in $\omega_{\bm{k}}$ at the Goldstone vectors $\bm{k}\in G$ proportional to $1/L^2$). Introducing\cite{luitz2015} $\Theta_{\bm{k}} =\sqrt{(A_{\bm{k}}-B_{\bm{k}})(A_{\bm{k}}+B_{\bm{k}})^{-1}}$, we can write
\begin{equation}
\lambda_{k_y} = \frac{1}{2L}\sqrt{\sum_{k_x}\Theta_{k_x,k_y} \sum_{k'_x} \Theta^{-1}_{k'_x,k_y}}-\frac{1}{2}.
\end{equation}
Using (\ref{AB}) and (\ref{h-leading}), it is seen that for large $L$, when $h\propto L^{-4}$ is very small, $\Theta_{\bm{0}}\simeq \Upsilon_{\bm{0}}\sqrt{S/h}$ and $\Theta_{\pm \bm{Q}}^{-1}\simeq \Upsilon_{\bm{Q}}\sqrt{S/h}$ are $\mathcal{O}(L^2)$, while other $\Theta_{\bm{k}}^{\pm 1}$ are $\mathcal{O}(L^0)$. This gives asymptotically ($L\to\infty$) 
\begin{subequations}
\begin{eqnarray}
\lambda_0 &\sim & \sqrt{\frac{\alpha_0 \Upsilon_{\bm{0}}m_{\rm{AF}}L}{\Upsilon_{\bm{0}}+(N_G-1)\Upsilon_{\bm{Q}}}} \label{lambda-0-approx}\\ \hfill \notag \\
\lambda_{\pm q} &\sim & \sqrt{\frac{\alpha_{q} \Upsilon_{\bm{Q}}m_{\rm{AF}}L}{\Upsilon_{\bm{0}}+(N_G-1)\Upsilon_{\bm{Q}}}} \label{lambda-Q-approx}
\end{eqnarray}
\label{large-occ-asymptotics}
\end{subequations}
where 
\begin{subequations}
\begin{eqnarray}
\alpha_0 &=& \frac{1}{2\pi}\int_{-\pi}^\pi \mathrm{d}k_x \Theta_{k_x,0}^{-1} \label{alpha-0-integral}\\
\alpha_{q} &=& \frac{1}{2\pi}\int_{-\pi}^\pi \mathrm{d}k_x \Theta_{k_x,q}
\label{alpha-Q-integral}
\end{eqnarray}
\label{alpha-integrals}
\end{subequations}
with the integrands in (\ref{alpha-integrals}) evaluated for $h=0$. The leading correction to (\ref{lambda-0-approx})-(\ref{lambda-Q-approx})  is the constant $-1/2$, with additional corrections of $O(L^{-1/2})$. We will refer to the $N_G$ $\mathcal{O}(L^{1/2})$ quantities $\lambda_0$ and $\lambda_{\pm q}$ as the large occupation numbers. To leading order in their inverses, their contribution to the R\'{e}nyi entropy (\ref{renyi-lambda}) is 
\begin{eqnarray}
\lefteqn{S_n^{\rm{large }\;\lambda} = \ln \lambda_0 + (N_G-1)\ln \lambda_{q} =} \nonumber \\ & & \hspace{-0.8cm} \frac{N_G}{2} \ln \left[\frac{(\alpha_0 \Upsilon_{\bm{0}})^{1/N_G}(\alpha_{q}\Upsilon_{\bm{Q}})^{1-1/N_G}}{\Upsilon_{\bm{0}} + (N_G-1)\Upsilon_{\bm{Q}}}m_{\rm{AF}} L \right]
\label{Sq-large}
\end{eqnarray}
which is independent of the R\'{e}nyi index $n$ and contains the universal logarithmic correction $(N_G/2)\ln L$ that signals the broken continuous symmetry in the thermodynamic limit. This suggests that $\lim_{L\to\infty}\lambda_{0,q}/\sqrt{L}$ may be regarded as a kind of alternative order parameter. This view is also consistent with the $\sqrt{m_{\rm{AF}}}$ factor in  (\ref{large-occ-asymptotics}). The main effect of the other $L-N_G$ occupation numbers on the R\'{e}nyi entropy is to give an area law term $(\propto L)$.\cite{luitz2015}

For the collinear phase ($N_G=2$), the large-$L$ asymptotic expressions simplify to
\begin{eqnarray}
\lambda_0 &=& \lambda_{q} \sim  \sqrt{\frac{\alpha_0 m_{\rm{AF}}L}{2}}, \label{lambda-0-approx-coll}\\
S_n^{\rm{large }\;\lambda} & = & \ln\left(\frac{\alpha_0  m_{\rm{AF}}L}{2}\right).
\label{Sn-large-lambda-approx-coll}
\end{eqnarray}


\section{Heisenberg quantum antiferromagnet on an anisotropic triangular lattice}
\label{results}

\subsection{Model}

In this section we present results for a Heisenberg quantum antiferromagnet with antiferromagnetic exchange interactions $J_1$ and $J'$, shown in Fig. \ref{square-triangle} (note that $J'$ involves only those diagonal bonds parallel to $y=+x$). The model has several interesting special cases (we set $J_1=1$ in the following): $J'=0$ and $J'=1$ correspond to the nearest-neighbor Heisenberg model on the square and triangular lattice, respectively, and $J'\to \infty$ to the limit of decoupled chains. For a general value of $J'$ the model is equivalent to a nearest-neighbor model on the triangular lattice with some bond anisotropy (two directions on the triangular lattice have $J_1$ bonds while the third has $J'$ bonds). Using 
\begin{equation}
J(\bm{k}) = 2[\cos k_x + \cos k_y + J' \cos(k_x + k_y) ]
\end{equation}
the ordering vector is $\bm{Q}=(q,q)$ with $q=\pi$ for $0\leq J' \leq 1/2$ (collinear order) and $q=\arccos(-1/2J')$ for $J'> 1/2$ (coplanar spiral order) in the classical model. The quantum version of the model with spin $S$ was studied with LSWT in Ref. \onlinecite{merino99}. By calculating the sublattice magnetization $m_{\rm{AF}}$ it was found that the collinearly ordered phase persists all the way down to $S=1/2$ 
for $0 \leq J' < 1/2$, while the phase with spiral order exists for $1/2<J'<J'_c(S)$, where $J'_c(S)$ decreases with decreasing $S$, with $J'_c(1/2)\approx 3.75$. For $J'>J'_c(S)$ the ground state is magnetically disordered. 

In the next subsections we present numerical results for the R\'{e}nyi entropy $S_n$ and the occupation numbers $\lambda_{k_y}$ in the collinear and spiral phases, and compare some of these results with the analytical approximations in Sec. \ref{large-L}. 
As we only consider commensurate 
magnetic order (cf. remarks after Eq. (\ref{h-eq-1})), we will in the spiral phase, where $q$ varies continuously with $J'$, treat $q$ and not $J'$ as the independent variable. In this phase the number of magnetic sublattices $s>3$ except for $J'=1$ which has $s=3$.

\subsection{R\'{e}nyi entropy}
\label{renyi-anisotropic}

We have computed R\'{e}nyi entropies from Eq. \eqref{renyi-lambda} for the $S=1/2$ model for subsystem sizes ranging from $L_{\rm{min}}=10^2$ to $L_{\rm{max}}=5\times 10^4$. The most general fitting ansatz we have considered for $S_n$ is\cite{song2011,luitz2015,laflorencie2015} 
\begin{eqnarray}
S_n &=& a_n L + b_n \ln L + c_n + d_n/L \nonumber \\ &+& e_n \ln L/L + f_n  \ln \ln L + g_n\ln\ln\ln L.
\label{scaling-form}
\end{eqnarray}
The dominant contribution is linear in $L$, which is the well-known ``area law''. Moreover, within a given ordered phase, the R\'{e}nyi entropy is found to have a subleading logarithmic correction whose prefactor $b_n$ is to good accuracy equal to the universal and $n$-independent value $N_G/2$. Because of this $n$-independence, we have also considered an alternative expression for Eq. (\ref{renyi-lambda}) valid for integer $n$, when the binomial series for $(1+\lambda_{k_y})^n$ is finite:
\begin{equation}
S_n = \sum_{k_y}\ln \lambda_{k_y} + \frac{1}{n-1}\sum_{k_y}\ln\left( \sum_{p=1}^n {{n}\choose{p}} \lambda_{k_y}^{1-p}\right). 
\label{q-indep}
\end{equation}
Here the $n$-independent part has been explicitly separated out as the first term. We have fitted minus this term to the scaling form (\ref{scaling-form}); the associated prefactor of the $\ln L$ term is referred to as $-b^*$.

\begin{figure}[h]
\centering
\includegraphics[width=0.23\textwidth]{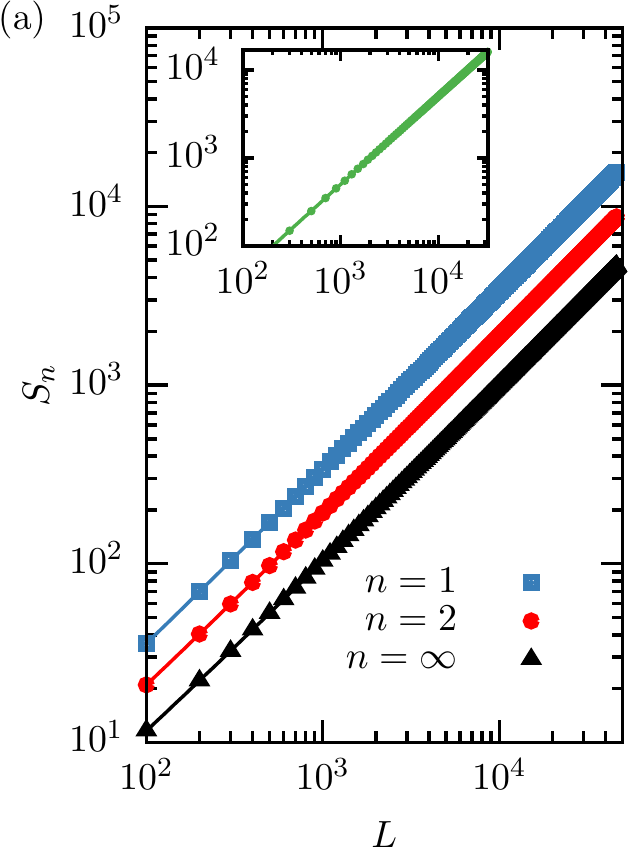}
\includegraphics[width=0.23\textwidth]{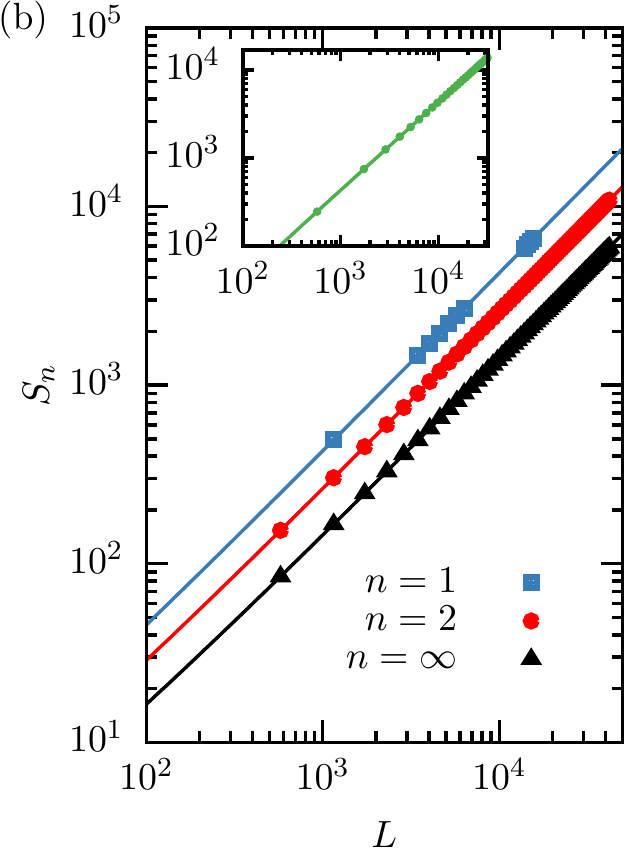}
\caption{
The R\'{e}nyi entropy $S_n(L)$ for the spin-$1/2$ Heisenberg model on the anisotropic triangular lattice, fitted to the ansatz $abcdf$ extracted from (\ref{scaling-form}) for $n=1$ (blue), $n=2$ (red), $n=\infty$ (black). (a) $J'=0.25$ (collinear phase). (b) $J'=1$ (spiral phase). The insets show minus the $n$-independent part of the entropy, $-\sum_{k_y} \ln \lambda_{k_y}$, fitted to the same scaling form.}
\label{renyi-collinear}
\end{figure}

\begin{table}[htbp]
  \begin{tabular}{|c|c|c|}
    \hline
    \hline
      Ansatz & $J'=0.25$  & $J'=1$ \\\hline
    $abc$      & $2\cdot 0.443$ &$3\cdot 0.436$       \\
    $abcd$      & $2\cdot 0.451$ &$3\cdot 0.449$     \\
    $abcde$ & $2\cdot 0.456$  &$3\cdot 0.455$       \\
    $abcdf$   &  $2\cdot 0.498$      &  $3\cdot 0.499 $                  \\
    $abcdg$      & $2\cdot 0.484$  &$3\cdot 0.484$       \\
    $abcdfg$      & $2\cdot 0.487$ &$3\cdot 0.482$   \\
    $abcdef$ & $2\cdot 0.496$ &$3\cdot 0.495$        \\
    $abcdeg$   &  $2\cdot 0.484$   &  $3\cdot 0.484$ \\
    $abcdefg$   &  $2\cdot 0.502$   &  $3\cdot 0.501$ \\
    \hline
    \hline
  \end{tabular}
  \caption{Numerical estimates of the prefactor $b_1$, as obtained from fits to R\'{e}nyi entropy data for the spin-$1/2$ Heisenberg model on the anisotropic triangular lattice. The ansatz column lists the coefficients included from (\ref{scaling-form}). The range of $L$ used is $ [1000,46000]$.}
  \label{fit-various}
\end{table}
Fitted results for $b_1$ for two points within the collinear and spiral phases are given in Table \ref{fit-various}. Among the various  ansatze considered, $abcdf$ stands out by being both relatively simple and giving an accurate value for $b_1$ (i.e. close to the theoretically expected value $N_G/2$). Fitted results for $b_n$ ($n=1,2,\infty$) and $b^*$ for the ansatz $abcdf$ for some selected points within the collinear and spiral phases are given in Table \ref{lq}. Except for $n=\infty$, the results are generally within $1 \%$ of $N_G/2$. Fig. \ref{renyi-collinear} displays the $S_n$ data and $abcdf$ fits for two of the points. 
\begin{table}[htbp]\centering
  \begin{tabular}{|c|cc|cc|}
    \hline
    \hline
     & Collinear phase & & Spiral phase &  \\
    \cline{2-5}
     $J'$ & $0$ & $0.25$ & $1/\sqrt{2}$ & $1$  \\
    \hline
    $b_1$      & $2\cdot 0.497$ &$2\cdot 0.498 $ &$3\cdot 0.500$ & $3\cdot 0.499$       \\
    $b_2$      & $2\cdot 0.495$ &$2\cdot 0.496$ &$3\cdot 0.498$ & $3\cdot 0.496$      \\
    $b_\infty$ & $2\cdot 0.521$ &$2\cdot 0.527$ &$3\cdot 0.537$ & $3\cdot 0.530$      \\
    $b^*$   &  $2\cdot 0.495$   &$2\cdot 0.496$ &$3\cdot 0.499$ & $3\cdot 0.496$                  \\
    \hline
    \hline
  \end{tabular}
  \caption{Numerical estimates of the prefactor of $\ln L$ for the ansatz $abcdf$ for $S_n$ (cf. (\ref{scaling-form})), as obtained from fits to R\'{e}nyi entropy data for the spin-$1/2$ Heisenberg model on the anisotropic triangular lattice. The estimates are close to the value $N_G/2$.}
\label{lq}
\end{table}

\subsection{Occupation numbers for modes of the entanglement Hamiltonian}

In this section we present numerical results for the occupation numbers $\lambda_{k_y}$ calculated from Eq. (\ref{lambdaky-1}). Comparisons with the analytical approximations in Sec. \ref{large-L} are also presented. 

\begin{figure*}
\centering
\includegraphics[width=0.4\textwidth]{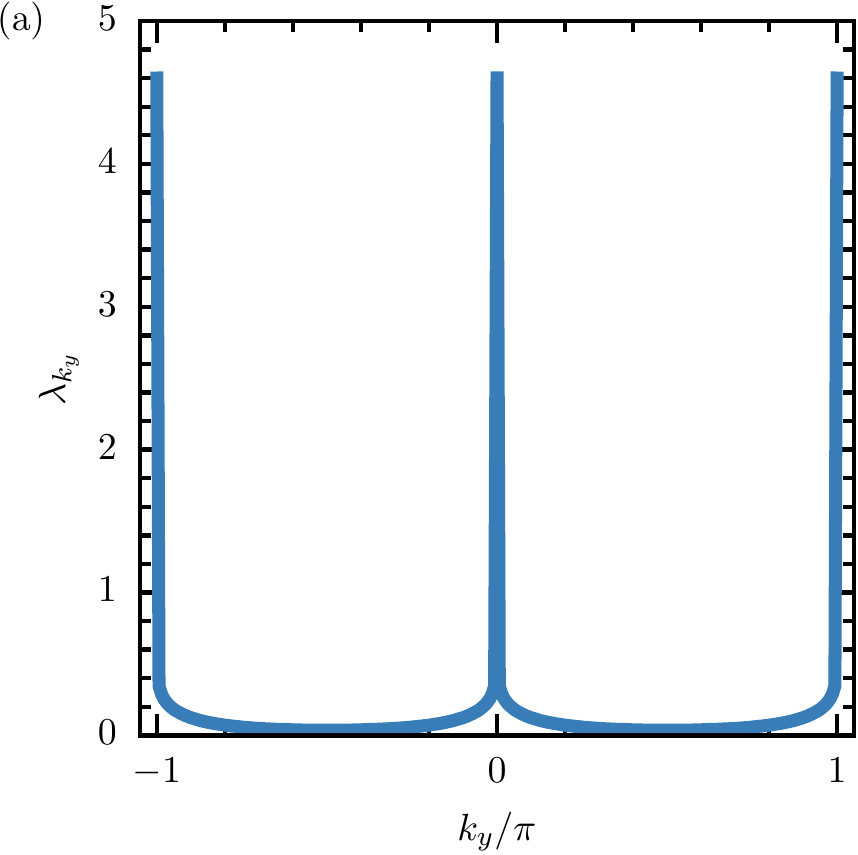}\hspace{1cm}
\includegraphics[width=0.4\textwidth]{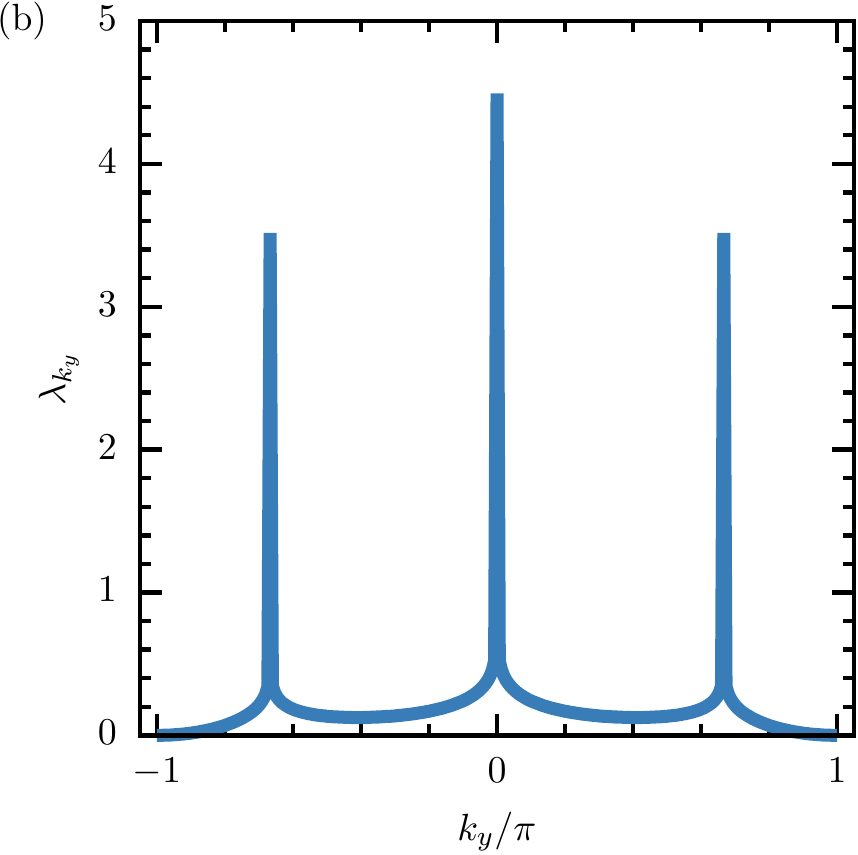}
\caption{The occupation number $\lambda_{k_y}$ of the mode $(x,k_y)$ of the entanglement Hamiltonian (\ref{H-E}) for the Heisenberg model on the anisotropic triangular lattice, for $S=1/2$ and $L=300$. (a)  $J'=0$ (collinear phase). (b)  $J'=1$ (spiral phase). There are $N_G$ inequivalent peaks due to $\lambda_0$ and $\lambda_{\pm q}$, where $N_G$, the number of Goldstone modes, equals 2 for the collinear phase and 3 for the spiral phase (note that for the collinear phase, $\pm q = \pm \pi$ are equivalent wave vectors). In the collinear phase the two peaks have the same height, while in the spiral phase 
the peaks at $\pm q$ are smaller than the peak at $0$.}
\label{lambda-ky}
\end{figure*}

Fig. \ref{lambda-ky} shows $\lambda_{k_y}$ for $-\pi\leq k_y \leq \pi$ (the equivalent wavevectors $\pm \pi$ are both included) for two values of $J'$ belonging to the collinear and spiral phase, respectively. There are sharp peaks for $k_y=0$ and $k_y=\pm q$. The number of inequivalent peaks is 2 in the collinear phase (since there $\pm q=\pm \pi$ are equivalent) and 3 in the noncollinear phase; in each phase this number equals $N_G$, the number of Goldstone modes. In the collinear phase the peaks at $0$ and $q$ have the same height (cf. Eq. (\ref{symm-collinear})), while in the spiral phase the peak at $0$ is larger than the equal-height peaks at $\pm q$. In comparison, the remaining $L-N_G$ occupation numbers are small and vary relatively little with $k_y$.

\begin{figure}[h!]
\centering

\includegraphics[width=0.45\textwidth]{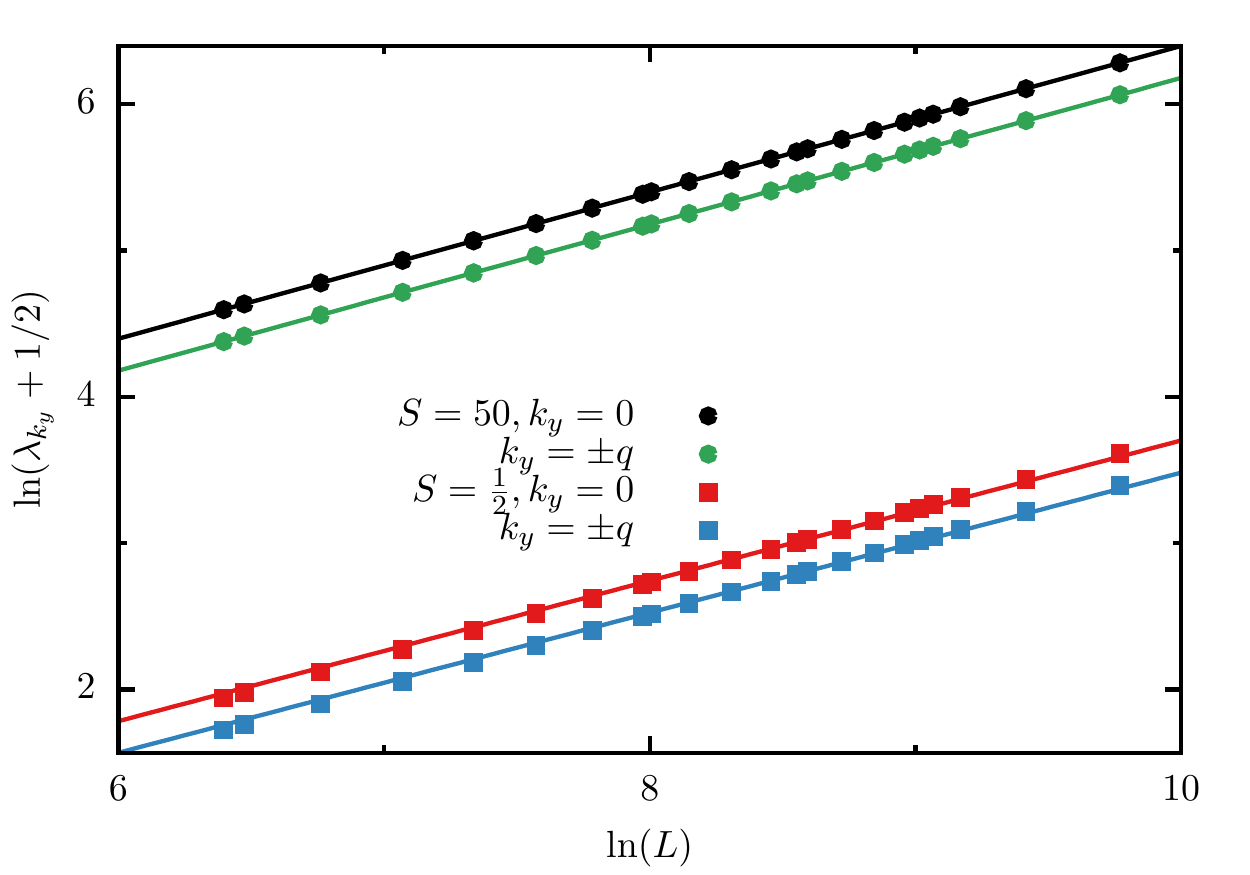}
\caption{Log-log plots of $\lambda_{0,q}+1/2$ versus $L$ in the spiral phase ($J'=1$) for $S=\frac{1}{2}$ and $S=50$. The full curves are fits to $p \ln L + \kappa$, with fitting parameters $p$ and $\kappa$  given in Table \ref{scaling-L}.}
\label{lambda-L1}
\end{figure}

To check the $L\to\infty$ asymptotic expressions 
 for the large occupation numbers derived in Sec. \ref{large-L}, we have fitted $\ln (\lambda_{0,q}+1/2)$ to the form $p \ln L + \kappa$ in both the collinear and spiral phase, for $S=1/2$ and $S=50$. Table \ref{scaling-L} shows fitted values of $p$ and $\kappa$, which are in good agreement 
with the theoretically expected value $p=1/2$ and the analytical expression for $\kappa$ (denoted by $\kappa_a$ in the table) obtained from Eqs. (\ref{lambda-0-approx})-(\ref{lambda-Q-approx}).  The associated plots for the spiral phase ($J'=1$) are shown in Fig. \ref{lambda-L1}. 


\begin{table}[htbp]\centering
  \begin{tabular}{|l|c|ccc|ccc|}
    \hline
    \hline
     $J'$ & $k_y$ &  & $S=1/2$ & & & $S=50$ &  \\

     	& & $p$ & $\kappa$  & $\kappa_a$ & $p$ & $\kappa$ & $\kappa_a$\\
        \hline
      $0$ & $0,q$  & $0.499$  & $-1.22$ & $-1.23$ & $0.500$ & $1.32$ & $1.32 $   \\ 
           \hline 
     $1$   & $0$& $0.498$ &$-1.25$ & $-1.27$  & $0.500$& $1.40$ & $1.40$  \\ 
     &$\pm q$  & $0.498$ &$-1.47$ & $-1.49$  & $0.500$ & $1.18$ & $1.18$    \\
    \hline
    \hline
  \end{tabular}
\caption{Fitting parameters $p$ and $\kappa$ for the fit of $\ln (\lambda_{0,q}+1/2)$ to the form $p \ln L + \kappa$ (see also Fig. \ref{lambda-L1}). The minimum value of $L$ used in the fits is 1000.} 
\label{scaling-L}
\end{table}

\begin{figure}
\centering
\includegraphics[width=0.45\textwidth]{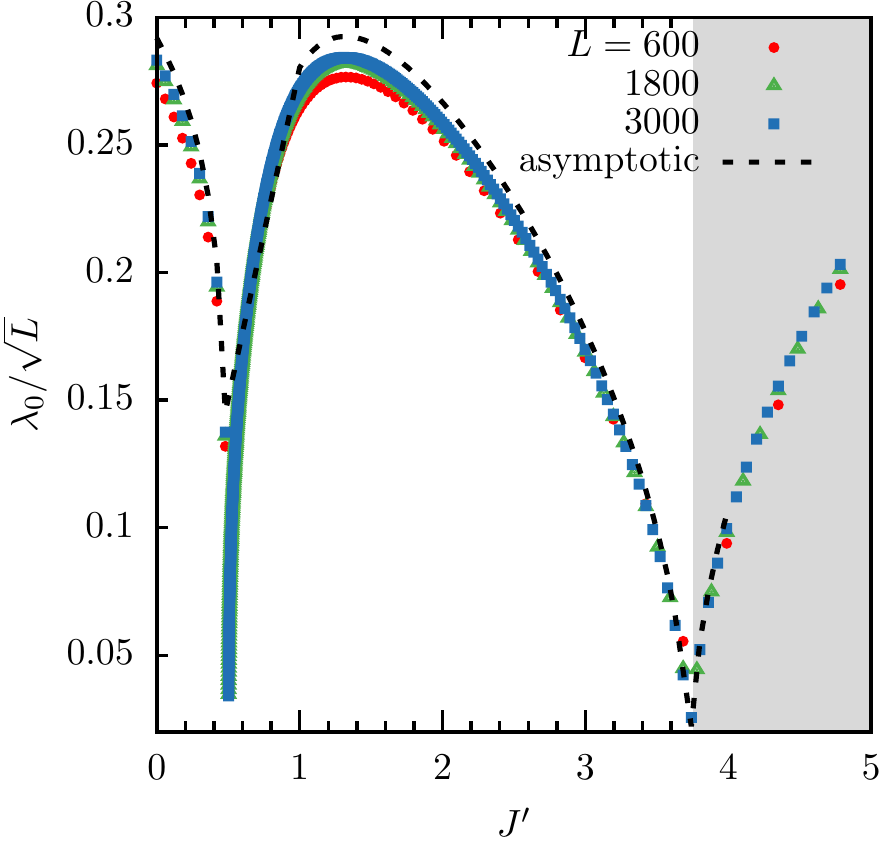}
\caption{$\lambda_0/\sqrt{L}$ as a function of $J'$ for $S=1/2$ and different values of $L$. As $L$ increases the curves approach the asymptotic value from  (\ref{lambda-0-approx}) (dashed line). The phase transitions at $J'=1/2$ and $J'\approx 3.75$ are visible as cusp-like endpoints of sharp dips. In the magnetically disordered phase (greyed-out region) there are no large occupation numbers, so $\lambda_0/\sqrt{L}$ should go to 0 there. The curves shown in this region are unphysical (see text for further details); they are included merely to make it easier to visually locate the phase transition. The plot illustrates that $\lim_{L\to\infty} \lambda_0/\sqrt{L}$ can be viewed as a kind of order parameter alternative to $m_{AF}$.}
\label{order-parameter}
\end{figure}

To explore the role of $\lim_{L\to\infty}\lambda_{0,q}/\sqrt{L}$ as a kind of alternative order parameters for the magnetically ordered phases, Fig. \ref{order-parameter} shows $\lambda_0/\sqrt{L}$ as a function of $J'$ for $S=1/2$, for various values of $L$ (a corresponding plot of $\lambda_q/\sqrt{L}$ shows qualitatively similar features). As $L$ is increased, $\lambda_0/\sqrt{L}$ approaches the asymptotic value from (\ref{lambda-0-approx}) (dashed curve). The evolution as $L\to\infty$ of the cusp-like endpoints of the sharp dips define phase transition points at $J'=1/2$ and $J'=J'_c(S)$, giving $J'_c(1/2) \approx 3.75$, in good agreement with the value obtained from the vanishing of $m_{\rm{AF}}$ (see Fig. 3 in Ref. \onlinecite{merino99}). We note that the uptick in the curves to the right of the spiral phase was obtained by continuing to use (\ref{h-leading-iteration}) for $h$ also beyond the point where $m^*$ passes through 0, even though LSWT becomes invalid then. We have chosen to include these curves since they make the identification of $J'_c$ easier, but as they do not correspond to any physical $\lambda_0$ (as $\sqrt{L}$-size occupation numbers do not exist in the disordered phase), we have greyed out the region $J'>J'_c$.

\begin{figure*}[t]
\centering
\includegraphics[width=0.4\textwidth]{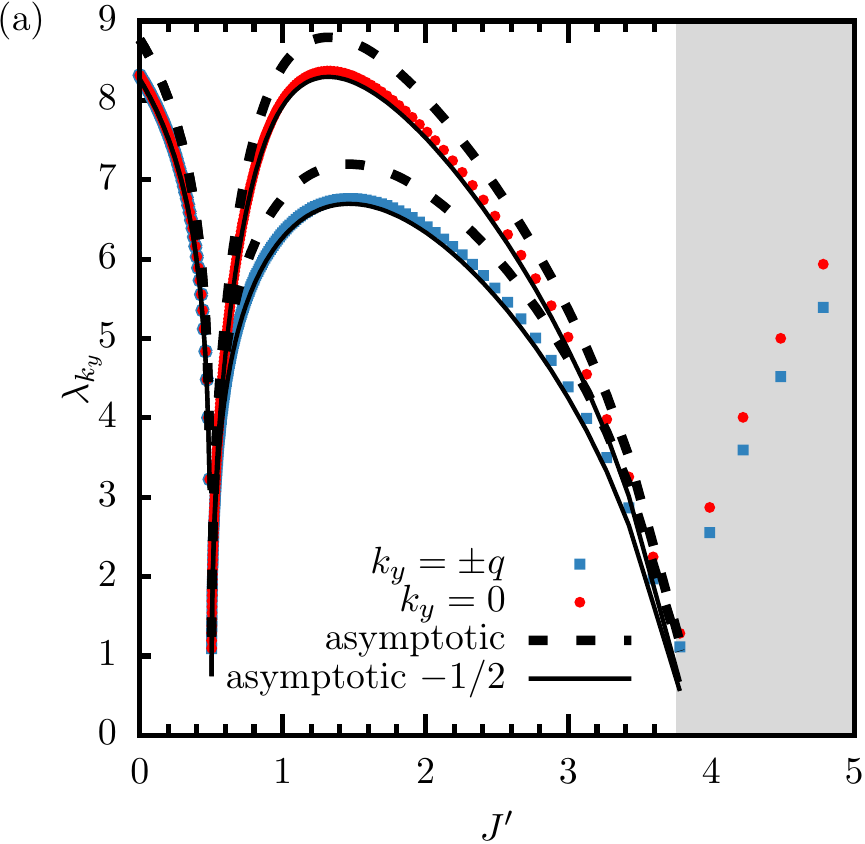}
\hspace{1cm}
\includegraphics[width=0.41\textwidth]{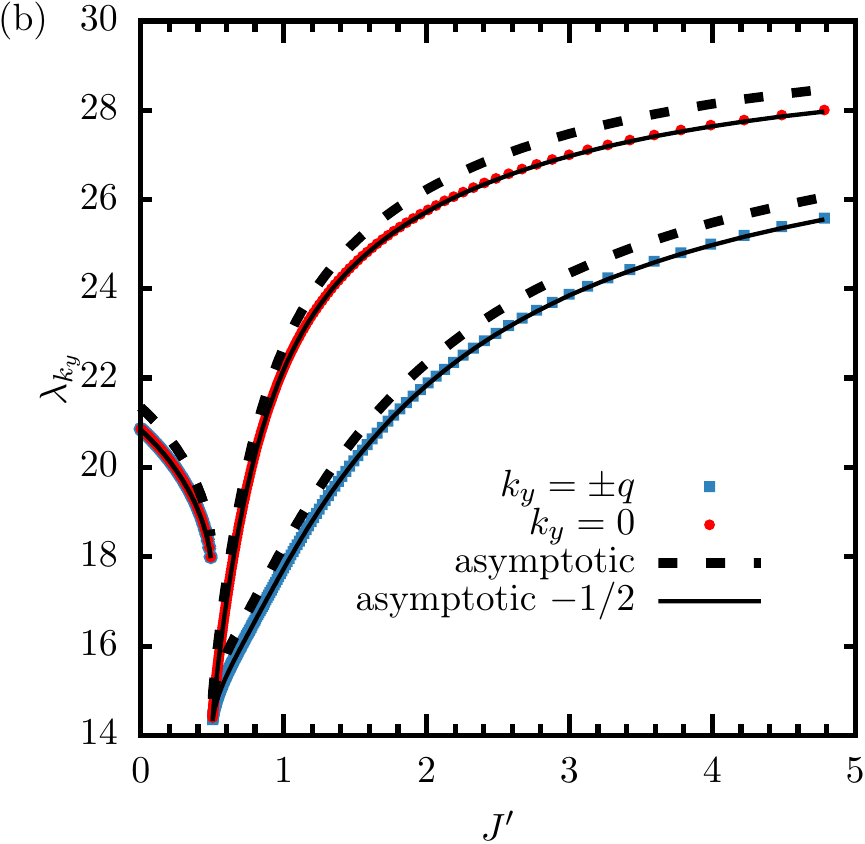}
\caption{The large occupation numbers $\lambda_{0}$, $\lambda_{\pm q}$ as functions of $J'$ for $L=900$. (a)  $S=\frac{1}{2}$. (b) $S=2$. The dashed lines show the large-$L$ asymptotic behavior (\ref{large-occ-asymptotics}), while the full lines also include the first subleading term $-1/2$. Whereas $\lambda_0=\lambda_q$ in the collinear phase ($J'<1/2$), $\lambda_0 >\lambda_q$ in the spiral phase ($1/2<J'< J'_c(S)$). The curves in the greyed-out region ($J'>J'_c(S)$) are unphysical (see text).}
\label{lambda-r}
\end{figure*}

A new feature of the spiral phase, not found in the collinear phase, is that $\lambda_0\neq \lambda_q$. (We will refer to this as ``anisotropy'' since it is partly related to differences in spin-wave velocities at $\bm{k}=\bm{0}$ and $\bm{k}=\pm\bm{Q}$.\cite{rademaker2015,chubukov1994}) This is seen in Fig. \ref{lambda-r} which shows $\lambda_0$ and $\lambda_q$ as a function of $J'$ for $S=1/2$ and 2. The  anisotropy goes away as $J'$ approaches the phase transition to the collinear phase at $J'=1/2$. The anisotropy also mostly goes away as $J'$ approaches the phase transition to a magnetically disordered phase at $J'=J'_c(S)$, as seen in the plot for $S=1/2$ (for $S=2$, $J'_c$ is well beyond the plot range for $J'$). In the analytic approximations (dashed and full lines in Fig. \ref{lambda-r}) the anisotropy in the spiral phase comes both from $\Upsilon_{\bm{0}} \neq \Upsilon_{\bm{Q}}$ and $\alpha_0\neq \alpha_q$. 

Fig. \ref{lambda-r} also shows that the value of $J'$ at which the large occupation numbers attain their maximum within the spiral phase depends on $S$ and is also different for $\lambda_0$ and $\lambda_{\pm q}$. Obviously the same will therefore hold for $\lambda_{0,q}/\sqrt{L}$. We note that, in contrast, the LSWT expression for $m_{\rm{AF}}$ attains its maximum at $J'=1$ independently of $S$.

\begin{figure}
\centering
\includegraphics[width=0.45\textwidth]{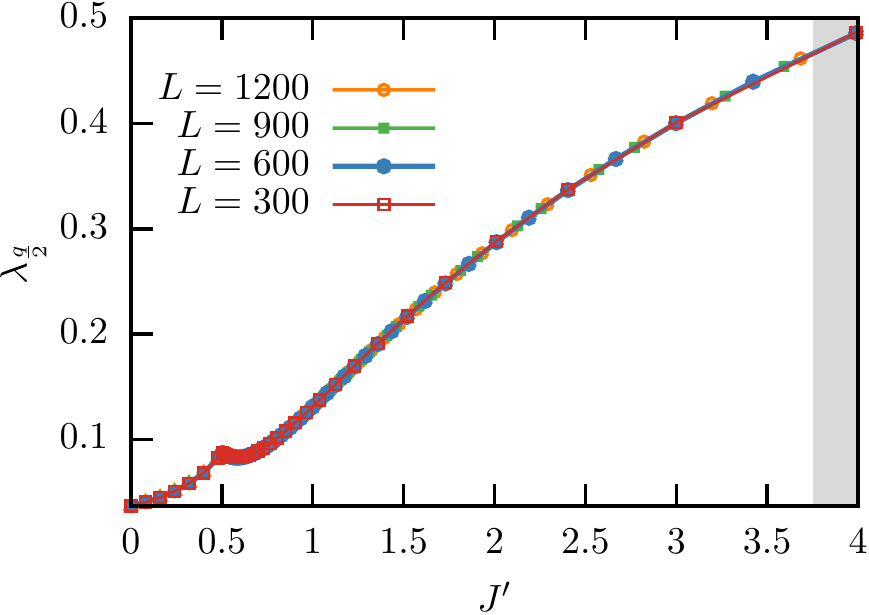}
\caption{$\lambda_{q/2}$ as an example of a ``small" occupation number, as a function of $J'$ for $S=1/2$ and various values of $L$. There is no visible $L$-dependence.  A cusp-like feature is observed at the phase transition at $J'=1/2$, while there is no sign of the phase transition to the disordered phase at $J'\approx 3.75$.}
\label{small-lambda-r}
\end{figure}

\begin{figure}
\centering
\includegraphics[width=0.45\textwidth]{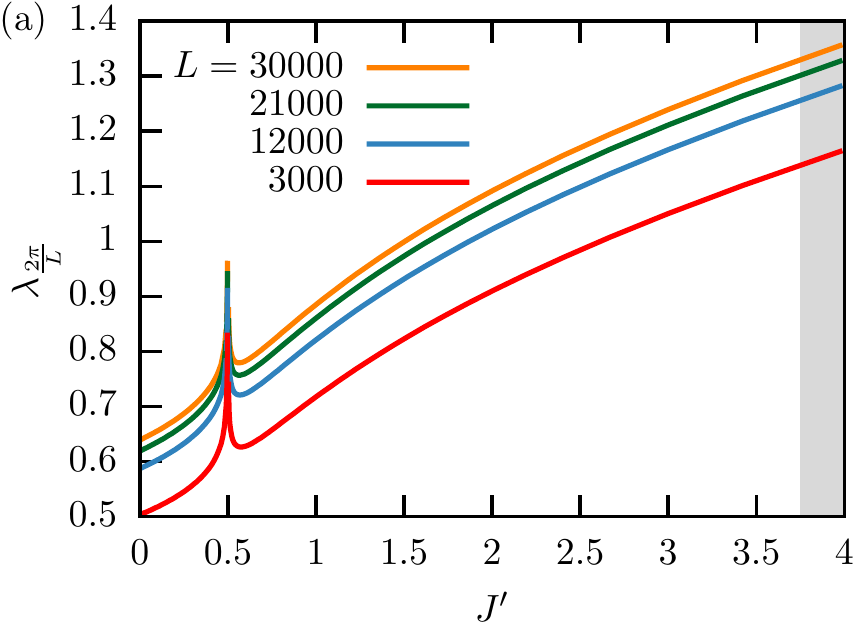}
\includegraphics[width=0.47\textwidth]{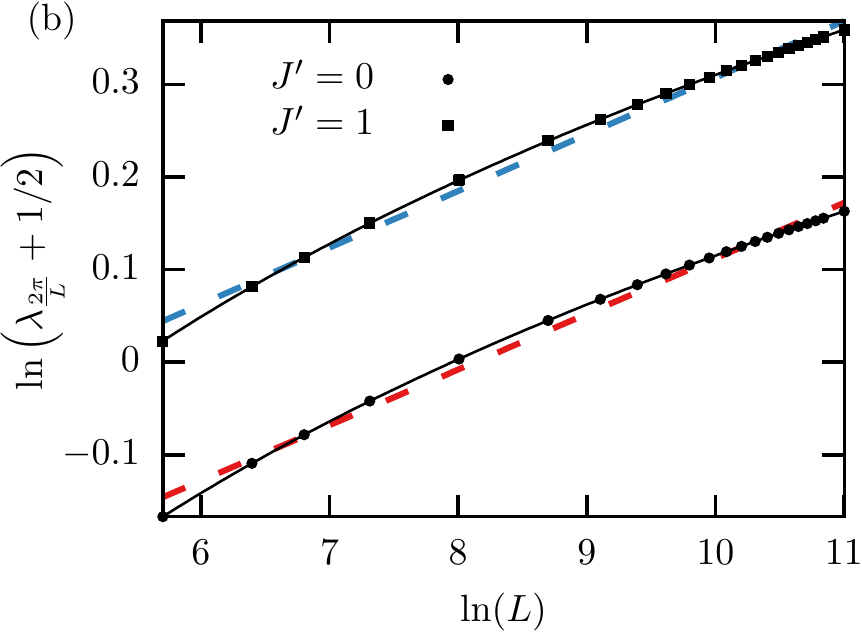}
\caption{(a) Occupation number for $k_y=2\pi/L$ (i.e. the nearest wavevector to $k_y=0$) as a function of $J'$ for $S=1/2$ and various values of $L$. A weak increase with $L$ is observed. A cusp-like feature is observed at the phase transition at $J'=1/2$, while there is no sign of the phase transition to the disordered phase at $J'\approx 3.75$ (grey region). (b) Fits of $\ln(\lambda_{2\pi/L}+1/2)$ to the ans\"{a}tze $p_1 \ln L + \kappa_1$ (dashed lines) and $p_2 \ln \ln L + \kappa_2$ (full lines). The former gives a poor fit and is also not expected for other reasons (see text). The latter gives excellent agreement, with $p_2\approx 0.5$ for both values of $J'$.}
\label{lambda-2piL}
\end{figure}

It remains to discuss the $L$-dependence of $\lambda_{k_y}$ for $k_y\neq 0,\pm q$. For most of these $k_y$ values there seems to be negligible $L$-dependence, as illustrated in Fig. \ref{small-lambda-r}, a behavior consistent with the area law term in $S_n$. The exception is for $k_y$ in very small regions around $0,\pm q$, for which the occupation numbers show a weak increase with $L$, as illustrated for $\lambda_{2\pi/L}$ in Fig. \ref{lambda-2piL}(a). The $L$-dependence is investigated in more detail in Fig. \ref{lambda-2piL}(b). If it were power-law  ($\lambda_{k_y}\propto L^{p_1}$ with $p_1>0$), the coefficient $b_n$ in $S_n$ would be larger than $N_G/2$. This is neither expected from the general theory, nor consistent with the numerical fits of $S_n$ in Sec. \ref{renyi-anisotropic}, and a power-law fit to $\lambda_{2\pi/L}$ indeed gives poor agreement. Instead excellent agrement is found to a fit to $\lambda_{2\pi/L}\propto (\ln L)^{p_2}$ with $p_2\approx 1/2$. 

This $L$-dependence of $\lambda_{2\pi/L}$ would give a term $p_2 \ln \ln L$ in $S_n$ and thus contributes to the coefficient $f_n$ in (\ref{scaling-form}), although numerically it appears that the latter also has other significant contributions. We note that connections between the presence of a $\ln \ln L$ term in $S_n$ and the $L$-dependence of modes close to those with lowest "energy" (and thus with the highest occupation) in the entanglement Hamiltonian have also been discussed for some spin-wave models in Ref. \onlinecite{frerot2015}, where the subsystem was however taken to be a half-torus ($L\times L/2$ sites). 

\section{Summary and discussion}
\label{discussion}

\subsection{Summary of results}

In this paper we have used MLSWT to study a class of square-lattice Heisenberg antiferromagnets whose exchange interactions have a Fourier transform $J(\bm{k})$ that we restrict to have at most two inequivalent minima located at $\pm\bm{Q}$ where $\bm{Q}=(q,q)$. Classically this gives magnetic order which for general $q$ is coplanar spiral, reducing to a collinear order if $q=\pi$ ($\pm \bm{Q}$ equivalent). When magnetic order survives in the semiclassical LSWT analysis, there are $N_G=3$ Goldstone modes (at $\bm{0},\pm\bm{Q}$) for spiral order, while $N_G=2$ for collinear order. Despite our square-lattice formulation, the theory can by appropriate choice of the exchange interactions also describe models most naturally defined on a triangular lattice. 

For a subsystem $\mathcal{A}$ of $L$ sites wrapping around the $y$ direction of an $L\times L$ torus, $S_n$  is expressed as a $k_y$-sum involving the occupation numbers $\lambda_{k_y}$ of the modes of the entanglement Hamiltonian. Because of the one-dimensionality and translational invariance of ${\cal A}$, several properties can be understood analytically for large $L$.  
The area law term in $S_n$ can be understood from the fact that $\sim L$ of the occupation numbers are $L$-independent.\cite{luitz2015}   
In contrast, the $N_G$ occupation numbers $\lambda_0$ and $\lambda_{\pm q}$ scale like $L^{1/2}$, which gives the universal and $n$-independent term $(N_G/2)\ln L$ in $S_n$. While this result was found from MLSWT for the collinear $q=\pi$ case with $N_G=2$ in Ref. \onlinecite{luitz2015}, our treatment extends it to a general $q$ commensurate with the lattice, i.e. $q=2\pi r/s$ with $r$ and $s$ coprime integers, $s$ being the number of magnetic sublattices, with $s=2$ for collinear order and $s\geq 3$ for spiral order. The coefficients of the $L^{1/2}$ scaling give a nonuniversal $n$-independent term in $S_n$. The coefficients of $\lambda_0$ and $\lambda_q$ are the same in the collinear case but different in the spiral case. We find analytical expressions for these coefficients, and suggest that they can be viewed as alternative order parameters. 

Sec. \ref{results} considered numerical calculations for a spin-$S$ Heisenberg antiferromagnet on an anisotropic triangular lattice for which the bonds along one of the three directions have exchange $J'$ instead of 1. This model has a collinear phase for $J'<1/2$, and a coplanar phase with $\bm{Q}$ continuously varying with $J'$ for $1/2<J'<J_c(S)$. Results consistent with the presence of a universal and $n$-independent term $(N_G/2)\ln L$ subleading to an area law term in $S_n$ were established for $S=1/2$ for selected values of $J'$ in the collinear and spiral phases, by curve fits to $S_n$ for $n=1$, 2, and $\infty$, and also to an expression for the $n$-independent part of $S_n$ valid for integer $n$. The most satisfactory fits contain a term $\propto \ln \ln L$ in the scaling ansatz for $S_n$, the need for which was also rationalized from analyzing the weak $L$-dependence of the occupation numbers $\lambda_{k_y}$ for $k_y$ in the vicinity of $0$, $\pm q$.\cite{frerot2015} We also explored, as functions of $J'$, the ''alternative order parameter'' $\lambda_0/\sqrt{L}$ and the anisotropy $\lambda_0\neq \lambda_q$ in the spiral phase.

\subsection{Discussion}

A central result of this work is the prediction of a universal term $(3/2)\ln L$ in $S_n$ for coplanarly ordered states with $N_G=3$ Goldstone modes. This was found for $3$-sublattice order ($q=2\pi/3$) with the NLSM approach in Ref. \onlinecite{rademaker2015}, and our MLSWT treatment generalizes it to $s$-sublattice order ($q=2\pi r/s$) for $s\geq 3$. Although this generalization has so far not been established by other methods, it fits with the expectation\cite{MG} that the $\ln L$-coefficient in two-dimensional systems should equal half the number of Goldstone modes. 

For $s=2$ and $s=3$ order it has been shown that the low-energy entanglement spectrum has the same structure as the TOS spectrum.\cite{MG,rademaker2015} To explore this
for $s>3$ order, it would seem ideal to identify a model for which an $s>3$ order with a rather small $s$ arises at the classical level for a rational and simple ratio of exchange constants, and for which this order survives as the spin $S$ is lowered all the way down to $1/2$. For the $J_1-J'$ model studied in Sec. \ref{results}, $s>3$ order occurs in the classical model only for rather special and irrational values of $J'/J_1$ (for generic values of $J'/J_1$ in the spiral phase the order is instead incommensurate). Furthermore, series expansions for the $S=1/2$ model\cite{weihong1999}  find that the dependence of the ordering wave vector on $J'/J_1$ in the spiral phase changes from the classical prediction (and also that the phase moves rightward and narrows a little compared to the LSWT prediction). A perhaps more suitable candidate is the nearest-neighbor Heisenberg model on the more complicated maple-leaf lattice, for which the 6-sublattice order of the classical model appears to survive also for $S=1/2$, as argued from exact diagonalization (including a TOS investigation), LSWT, and a variational method.\cite{schulenburg2000,schmalfuss2002}

Our MLSWT analysis could be extended in several directions. One could consider subsystems of a more general shape and size, as this would allow exploring other kinds of contributions to $S_n$, e.g. the potential existence of an analogy for coplanar order to the universal function $\gamma_n^{\rm{ord}}$ for collinear order,\cite{MG,laflorencie2015} contributions due to subsystem corners etc. Compared to the line subsystem used here, some of the transparency would however be lost, as the correlation matrix could no longer be diagonalized analytically. One could also relax our assumptions in Sec. \ref{theory} concerning the number and locations of the minima of $J(\bm{k})$.

We conclude with some critical remarks on MLSWT. In spin-wave theory the expansion "parameter" is $n_j/2S$. LSWT gives $\langle n_j\rangle$ independent of $S$, so (provided $\langle n_j\rangle$ doesn't diverge) $\langle n_j\rangle/2S$ can be made arbitrarily small by increasing $S$,  which justifies ignoring terms in $H$ of higher order in $n_j/2S$ for large $S$. In contrast, in MLSWT the condition of vanishing sublattice magnetization implies $\langle n_j\rangle/2S = 1/2$, which makes the neglect of higher-order terms harder to justify, regardless of how large $S$ is. Also, MLSWT is not able to reproduce the spin-rotation invariance of the spin-spin correlations of the true finite-size ground state.\cite{song2011} 
On a related note, Ref. \onlinecite{frerot2015} argued that in MLSWT the symmetry is "broken "twice"" rather than being restored, and furthermore pointed out that the TOS spectrum is not correctly reproduced in MLSWT.  

In view of these criticisms, it may seem quite remarkable that MLSWT describes the R\'{e}nyi entropy of finite-size Heisenberg antiferromagnets rather well. Similar sentiments have previously been expressed in Refs. \onlinecite{song2011} and \onlinecite{frerot2015}. We hope that future work will shed more light on this issue.

\section*{Acknowledgments}

We acknowledge enlightening discussions with Louk Rademaker and Huan-Qiang Zhou. This work was partly supported by the Research Council of Norway through its Centres of Excellence funding scheme, project number 262633, ''QuSpin''.


\appendix

\bibliographystyle{unsrt}
\bibliography{mybib.bib}

\begin{references}
\bibitem{v2} We also note that in v2 of Ref. \onlinecite{MG} the analysis of the R\'{e}nyi entropy of the O(N) NLSM was strengthened with a ground state calculation valid for general $N$.
\bibitem{inverse-transforms} The inverse transformations are given by $\bm{S}_j = \frac{1}{\sqrt{N}}\sum_{\bm{k}} \bm{S}(\bm{k}) e^{i\bm{k}\cdot \bm{r}_j}$ and $J_{ij} = \frac{1}{N}\sum_{\bm{k}}J(\bm{k})e^{i\bm{k}\cdot(\bm{r}_i-\bm{r}_j)}$.
\bibitem{proof-symm-collinear} The $k_x$-sums in (\ref{lambdaky-1}) can be shifted by an arbitrary wavevector $\delta_x$. Eq. (\ref{symm-collinear}) follows from choosing the shift to be $-Q_x$ and using that in a collinear phase, $A_{\bm{k}+\bm{Q}}=A_{\bm{k}}$ and $B_{\bm{k}+\bm{Q}}=-B_{\bm{k}}$.
\end{references}

\end{document}